\documentclass[11pt]{article}
\usepackage{mathptmx}
\usepackage{sprinconf}
\usepackage{graphicx}
\usepackage{bm}
\usepackage{amsmath,amssymb}
\usepackage{graphics,hyperref,textcomp}

\begin{document}
\title{The Effect of Reduced Spatial Symmetries on Lattice States:  Results for Non-zero Linear Momentum}

\author{David C. Moore, George T. Fleming}

\affil{Sloane Physics Laboratory, Yale University, New Haven, CT
06520, USA}
\vspace{-36pt}                          

\beginabstract
The irreducible representations of the cubic space group are
described and used to determine the mapping of continuum states to
lattice states with non-zero linear momentum. The Clebsch-Gordan
decomposition is calculated from the character table for the cubic
space group. These results are used to identify multiparticle states
which appear in the hadron spectrum on the lattice.
\endabstract

\section{Introduction}
A question of great interest in lattice QCD is how rotational states
on the lattice correspond to states of definite angular momentum in
the continuum limit.  This problem has been discussed in several
contexts previously -- in solid state physics, the ``cubic
harmonics" are formed by a projection of the continuum spherical
harmonics onto the lattice, e.g.  \cite{Altmann65}.  In lattice QCD,
the reduction of continuum states to the hypercubic lattice was
given by Mandula \textit{et al.} \cite{Mandula83}, and the reduction
of the full continuum symmetry group (including Poincar\'{e}, color,
flavor, and baryon number symmetries) to the lattice for the case of
staggered fermions was given by Golterman and Smit
\cite{Golterman84a,Golterman84b} and expanded to include non-zero
momentum by Kilcup and Sharpe \cite{Kilcup86}.

We are interested in the classes of lattice actions with unbroken
flavor symmetries (i.e. Wilson and overlap).  In particular, we
focus on the IRs of the symmetry group of the lattice Hamiltonian.
Johnson \cite{Johnson82} considered the mapping of continuum SU(2)
states to the octahedral group and its double cover, and Basak
\textit{et al.} \cite{Basak05,Basak05b} considered the inclusion of
parity in these groups. We expand this work to include states at
non-zero momentum \cite{Moore:2005dw}.

In addition, we consider multiparticle states on the lattice by
calculating the Clebsch-Gordan decomposition of direct products of
lattice IRs. In many lattice calculations, the operators used to
compute correlation functions are constructed to transform
irreducibly under the symmetry group of continuum QCD Hamiltonian.
As above, it is well known that these operators need not transform
irreducibly under the symmetry group of the lattice QCD Hamiltonian.
Thus, it is important to understand how such continuum operators
transform on the lattice in order to correctly determine what
multiparticle states will appear in the lattice spectrum.  When
calculating ground state masses, ignoring this fact usually does not
lead to confusion. One possible exception is the $I(J^P)$ =
$\frac{1}{2}(\frac{3}{2}^+)$ and $\frac{1}{2}(\frac{5}{2}^+)$
baryons, whose lowest-lying resonances should correspond to the
experimentally observed $N(1720)\ P_{13}$ and $N(1680)\ F_{15}$,
respectively \cite{Basak:2004hr}.

We demonstrate for each set of quantum numbers in the center-of-mass
frame what two-particle decompositions are possible, including
states with non-zero relative momentum.

\section{The symmetry groups and their representations}
The continuum spacetime symmetries of the QCD action are given by
the Poincar\'{e} group, $\mathcal{P}$.  The spatial symmetries of
the QCD Hamiltonian correspond to a subgroup of $\mathcal{P}$ which
is the semidirect product of the group of orthogonal transformations
in three dimensions and the group of translations, $\mathcal{T}^3
\rtimes \mathrm{O}(3)$ (pure time inversion is also a symmetry
element, but its addition is trivial since it commutes with all
other group elements). On the lattice, the symmetry group is
$\mathcal{T}^3_{lat}\rtimes \mathrm{O}_h$, where the rotational
group is reduced to a subgroup of O(3) with only a finite number of
rotations and reflections, and where the subgroup of
$\mathcal{T}^3_{lat}$ contains only lattice translations. We
consider the double covers of the rotational groups since the double
valued IRs correspond to fermionic states in the continuum.

The group of proper rotations of a cube in three dimensions is the
octahedral group, O. We are also interested in its double cover,
$\mathrm{O^D}$, and we consider the inclusion of parity by forming
the group $\mathrm{O}^\mathrm{D}_h = \mathrm{O^D} \times
\mathrm{C_2}$, where $\mathrm{C_2}$ consists of the identity and a
parity element $I_s$, corresponding to inversion of all three
coordinate axes through the origin. The inclusion of parity is
straightforward because $I_s$ commutes with all proper rotations in
three dimensions.

For $\mathcal{T}^3_{lat}\rtimes \mathrm{O}^\mathrm{D}_h$, we can
then write the group elements $\{R_i, \mathbf{n}\}$, where $R_i$
denotes a rotation in one of the lattice rotation groups discussed
above, followed by a lattice translation by $\mathbf{n}$.  The
subgroup of translations, $\mathcal{T}^3_{lat}$, is normal, so we
can easily use this subgroup to induce the IRs of the full group. We
can construct the characters for the IRs of the continuous group
analogously. We then subduce a representation of the cubic space
group by considering the IRs of the continuum group restricted to
the subgroup of elements which are in the lattice group. By the
orthogonality properties of characters for finite groups, we can
decompose the subduced continuum representation into lattice IRs.
However, the group $\mathcal{T}^3_{lat}\rtimes
\mathrm{O}^\mathrm{D}_h$ allows arbitrarily large translations, so
we must consider the 3-torus formed by the boundary conditions
$\mathbf{r} + \mathbf{N} = \mathbf{r}$ for all vectors $\mathbf{r}$
and some constant vector $\mathbf{N} = (N,N,N)$.

For finite lattices, we find that the projection formula reduces to
the projection of the continuous rotation group to the little group
given by $\mathbf{k}$, independent of the lattice size. We also find
that representations labeled by different stars are orthogonal.
Therefore, the reduction of an arbitrary continuum IR labeled by
$(\mathbf{k}, m_j)$ contains IRs of the discrete group labeled by
$\mathbf{k}$, and by $\alpha$ which correspond to the reduction of
$\mathrm{O}^\mathrm{D}$(2) to the little group \cite{Moore:2005dw}.

If $\mathbf{k}=0$, one reduces $\mathrm{O}^\mathrm{D}$(3) to
$\mathrm{O}^\mathrm{D}_h$. These results can be read off those given
by Johnson \cite{Johnson82} using the result that IRs of
$\mathrm{O}^\mathrm{D}$(3) with positive parity, $\pi = +1$,
correspond to the ``gerade" IRs (\textit{e.g.}\  $A_{1g}$) of
$\mathrm{O}^\mathrm{D}_h$ only, and those with $\pi =-1$ correspond
to the ``ungerade" IRs (\textit{e.g.}\  $A_{1u}$) only .
\begin{table*}
\caption{\label{tab:red}Reduction of the double cover of O(2) to the
possible little groups.}
\begin{tabular}{lccccc}
  $m_j$ & $\mathrm{Dic}_4$ & $\mathrm{Dic}_3$ & $\mathrm{Dic}_2$ & $\mathrm{C_{4}}$ & $\mathrm{C_{2}}$ \\
  \hline
  $0^+$         & $A_1$            & $A_1$            & $A_1$            & $A_1$ & $A$\\
  $0^-$         & $A_2$            & $A_2$            & $A_2$            & $A_2$ & $A$\\
  $\frac{1}{2}$ & $E_1$            & $E_1$            & $E$              & $E$ & $2B$\\
  $1$           & $E_2$            & $E_2$            & $B_1 \oplus B_2$ & $A_1 \oplus A_2 $ & $2A$ \\
  $\frac{3}{2}$ & $E_3$            & $B_1 \oplus B_2$ & $E$              & $E$ & $2B$\\
  $2$           & $B_1 \oplus B_2$ & $E_2$            & $A_1 \oplus A_2$ & $A_1 \oplus A_2$ & $2A$\\
  $\frac{5}{2}$ & $E_3$            & $E_1$            & $E$              & $E$ & $2B$\\
  $3$           & $E_2$            & $A_1 \oplus A_2$ & $B_1 \oplus B_2$ & $A_1 \oplus A_2$ & $2A$\\
  $\frac{7}{2}$ & $E_1$            & $E_1$            & $E$              & $E$ & $2B$\\
  $4$           & $A_1 \oplus A_2$ & $E_2$            & $A_1 \oplus A_2$ & $A_1 \oplus A_2$ & $2A$\\
\end{tabular}
\end{table*}

\section{Clebsch-Gordan Decomposition}

Additionally, we wish to calculate the decomposition of the direct
product of IRs of $\mathcal{T}^3_{lat}\rtimes
\mathrm{O}^\mathrm{D}_h$ into a direct sum of IRs.  Using the
character table for $\mathcal{T}^3_{lat}\rtimes
\mathrm{O}^\mathrm{D}_h$, the character of a group element $g \in
\mathcal{T}^3_{lat}\rtimes \mathrm{O}^\mathrm{D}_h$ in the direct
product representation $\Gamma_i \otimes \Gamma_j$ is given as
$\chi^{\Gamma_i,\Gamma_j}(g) = \chi^{\Gamma_i}(g)
\chi^{\Gamma_j}(g)$. The direct product representations are then
decomposed into lattice IRs using their orthogonality properties.

As we expect, we see that linear momentum is conserved, i.e. the
product of two representations with momenta $|\mathbf{k_1}|$ and
$|\mathbf{k_2}|$ gives only representations labeled by
$|\mathbf{k}|$ which are the sum of some vector in the star of
$\mathbf{k_1}$ and some vector in the star of $\mathbf{k_2}$. Thus,
the direct product of two IRs of $\mathcal{T}^3_{lat} \rtimes
\mathrm{O}^\mathrm{D}_h$ contain IRs labeled by $|\mathbf{k}| = 0$
(the IRs of $\mathrm{O}^\mathrm{D}_h$) if and only if
$|\mathbf{k_1}| = |\mathbf{k_2}|$.  Complete tables of the
Clebsch-Gordan decomposition for each of the possible lattice
momenta are given in \cite{Moore2006:inprep}.

\section{Multiparticle States}
Since continuum IRs with $J \leq \frac{3}{2}$ remain irreducible
under the reduced symmetries of the lattice, the continuum relations
are recovered for low spins. For example, the $\pi$ with $J^P = 0^-$
lies in the irreducible representation $A_{1u}$, and the $\pi\pi$
multiparticle state lies in $A_{1u} \otimes A_{1u} = A_{1g}$, which
as expected corresponds to $J^P = 0^+$.

Higher spin states are less straightforward since multiple lattice
IRs appear for each spin. A spin 2 continuum state could lie in
either the $E$ or $T_2$ representations on the lattice, which leads
to different possibilities for multiparticle states depending on the
particular decomposition of the spin 2 continuum IR into lattice
IRs.

When combining states with non-zero momentum, the continuum
relations are not as easily recovered.  Here, the continuum
representations are no longer labeled by $J$, but by the projection
of $J$ along the momentum vector, $m_j$.  However, we know that a
particle with a given $m_j$ has $J \geq m_j$, so the lowest spin
state for a given irreducible representation of
$\mathcal{T}^3_{lat}\rtimes \mathrm{O}^\mathrm{D}_h$ will be $J =
m_j$. In addition, the reduced symmetry of the little groups leads
to fewer distinct lattice IRs than at zero momentum.  Since the
continuum IRs are mapped to fewer lattice representations, it is
more difficult to assign a particular spin to a given lattice
irreducible representation.

Thus, as we go to higher spins or non-zero momentum deviations will
occur from the continuum behavior.  For example, a $J^P = 2^+$ state
in the continuum can lie in either $E_g$, $T_{2g}$, or some
combination of both on the lattice. In the continuum, an $f_2$(1270)
meson with $J^P = 2^+$ has the decay modes $\pi\pi$, $4\pi$, and
$K\bar{K}$ \cite{PDBook}, and we expect to see multiparticle states
in the lattice spectrum corresponding to these decay modes. Indeed,
the multiparticle state $\{n,0,0\}; A_2 \otimes \{n,0,0\}; A_2$
corresponding to these decay modes occurs in the $E_g$ channel, but
not in the $T_{2g}$ channel. We must calculate exactly how the
particular spin 2 continuum state we are interested in subduces to
the lattice to determine whether these multiparticle states will
appear. In this case, it is possible that states we expect from the
continuum rules for addition of angular momentum would be absent
from the lattice spectrum.  Additionally, since each lattice IR
corresponds to many continuum states, the multiparticle states
expected for each of the corresponding continuum states appears in
the lattice channel. This leads to additional states in the lattice
spectrum.

\section{Conclusion}
We have found the mapping of continuum IRs to lattice states by
decomposing subduced continuum representations into the discrete
lattice IRs. In general, we find that a single continuum IR
decomposes into multiple IRs under the reduced symmetry of the
lattice.

These results were used to calculate the Clebsch-Gordan
decomposition and determine what multiparticle states can appear on
the lattice.  For low spins and zero momentum, we recover the
continuum behavior.  However, as we go to higher spins or non-zero
momentum, it is important to understand how our operators transform
irreducibly on the lattice since deviations from the continuum
behavior can occur.


\begin{thebibliography}{10}

\bibitem{Altmann65}
S.L. Altmann and A.P. Cracknell. {\em Rev. Mod. Phys.}, 37:19, 1965.

\bibitem{Mandula83}
J. Mandula, G. Zweig, and J. Govaerts. {\em Nucl. Phys.}, B228:91,
1983.

\bibitem{Golterman84a}
M. Golterman and J. Smit. {\em Nucl. Phys.}, B255:328, 1985.

\bibitem{Golterman84b}
M. Golterman and J. Smit. {\em Nucl. Phys.}, B245:61, 1984.

\bibitem{Kilcup86}
G.~W. Kilcup and S.~R. Sharpe. {\em Nucl. Phys.}, B283:493, 1987.

\bibitem{Johnson82}
R.~C. Johnson. {\em Phys. Lett.}, B114:147, 1982.

\bibitem{Basak05}
S.~Basak et~al. {\em Phys. Rev.}, D72:094506, 2005.

\bibitem{Basak05b}
S. Basak et~al. {\em Phys. Rev.}, D72:074501, 2005.

\bibitem{Moore:2005dw}
D. C. Moore and G. T. Fleming. {\em Phys. Rev.}, D73:014504, 2006.

\bibitem{Basak:2004hr}
S.~Basak et~al. {\em Nucl. Phys. Proc. Suppl.}, 140:278--280, 2005.

\bibitem{Moore2006:inprep}
D. C. Moore and G. T. Fleming, {\em in publ.} (hep-lat/0607004)

\bibitem{PDBook}
S.~{Eidelman}, et~al. {\em {Physics Letters B}}, 592:1+, 2004.


\end{thebibliography}
\end{document}